\documentclass[%
reprint,
superscriptaddress,
groupedaddress,
nofootinbib,
nobibnotes,
 amsmath,amssymb,
 aps,
prl,
]{revtex4-2}
\usepackage[dvipsnames]{xcolor}
\usepackage{graphicx}
\usepackage{dcolumn}
\usepackage{bm}
\usepackage[breaklinks=true,colorlinks,citecolor=blue,linkcolor=red,urlcolor=blue]{hyperref}

\usepackage{braket}
\usepackage[utf8]{inputenc} 
\usepackage{mathtools}
\usepackage[bottom]{footmisc}
\usepackage{hyperref}
\usepackage{amsfonts}
\usepackage{amsmath}
\usepackage{slashed}
\usepackage{amsthm}
\usepackage{mathbbol}
\usepackage{soul}

\usepackage{tikz}

\usepackage{tikz}

\begin{document}


\title{Inter-Dirac-Cone Coherence Beyond the Valley Framework}

\author{Yi-Chun Hung$\,^{\hyperlink{email1}{\dagger}}$}

\author{Arun Bansil$\,^{\hyperlink{email2}{\ddagger}}$}

\affiliation{Department of Physics,\;Northeastern\;University,\;Boston,\;Massachusetts\;02115,\;USA}
\affiliation{Quantum Materials and Sensing Institute,\;Northeastern University,\;Burlington,\;Massachusetts\;01803,\;USA}


\begin{abstract}
    
Although interactions are known to generate exotic phases in pseudospin-1/2 flat-band Dirac materials, it remains an open question whether higher-pseudospin Dirac cones continue to govern low-energy interaction-driven orders even without a well-defined valley degree of freedom. Here, we demonstrate a new form of inter-Dirac-cone coherence beyond the valley framework, which is realized through a staggered virtual-loop-current (SVLC) order in a partially filled pseudospin-1 Dirac flat band. Remarkably, the SVLC order continues to be driven by coherence between the pseudospin-1 Dirac cones even after the conventional valley-based description ceases to apply. The virtual loop currents are shown to originate from interaction-driven quantum fluctuations of charge densities and are found to exhibit an alternating circulation between the neighboring triangular plaquettes. The resulting spontaneous time-reversal symmetry breaking induces finite intrinsic anomalous Hall conductivity and orbital magnetization. The SVLC order is shown to be the lowest-energy solution in restricted real-space Hartree-Fock calculations in the weakly interacting regime. Our study establishes inter-Dirac-cone coherence as the organizing principle for interaction-driven orders in higher-pseudospin Dirac systems and identifies these systems as a new platform for driving exotic emergent phases beyond the valley framework.
\end{abstract}

\maketitle
\renewcommand{\thefootnote}{\fnsymbol{footnote}}
\footnotetext[2]{\hypertarget{email1}{Contact author: \href{mailto:hung.yi@northeastern.edu}{hung.yi@northeastern.edu}}}
\footnotetext[3]{\hypertarget{email2}{Contact author: \href{mailto:ar.bansil@northeastern.edu}{ar.bansil@northeastern.edu}}}
\par Flat bands in twisted bilayer graphene (TBG) and moiré transition-metal dichalcogenides (TMDs) provide a fertile platform for the emergence of exotic interaction-driven orders. The orbital features inherited from the underlying pseudospin-1/2 Dirac cones, massless in TBG and massive in moiré TMDs, stabilize symmetry-breaking orders such as intervalley coherence (IVC) \cite{PhysRevX.8.031089, PhysRevB.102.035136, PhysRevX.11.041063, Kim2023, Sun2021, Nuckolls2023, He2021, Devakul2021, Tuo_2025, Mu_oz_Segovia_2025, PhysRevB.104.075150, PhysRevB.110.045115}. Bipartite lattices provide an alternative to moiré engineering for flat bands, where lattice connectivity enforces flat bands via destructive interference \cite{10.1098/rspa.1955.0200, PhysRevLett.62.1201, Calugaru2022}, so that these flat bands remain robust even in the associated twisted bilayers \cite{PhysRevLett.133.236401, zhou2026, PhysRevB.109.155159}. Originating from the same lattice connectivity, certain bipartite lattices further host pseudospin-1 Dirac cones at the Brillouin zone (BZ) corner, as exemplified by the dice lattice \cite{10.1063/1.1666703, PhysRevB.34.5208}. The additional zero-energy flat-band states away from the pseudospin-1 Dirac cones, however, challenge the conventional valley-based description of the corresponding interaction-driven orders established for pseudospin-1/2 Dirac systems. 

\par However, flat bands in bipartite lattices provide a tunable platform for investigating interaction-driven orders, including magnetism \cite{10.1098/rspa.1955.0200, PhysRevLett.62.1201, doi:10.7566/JPSJ.86.063702, PhysRevB.111.024416} and superconductivity \cite{PhysRevResearch.5.043215, PhysRevB.70.134507, Mohanta2023}. These systems have also enabled studies of quantum geometric effects \cite{PhysRevB.95.024515, Wu2021} as well as interaction-driven topological phase transitions into high-Chern-number flat bands \cite{PhysRevB.84.241103, PhysRevB.104.235115}. Introducing a net flux can generate an all-flat-band system via Aharonov-Bohm cages \cite{PhysRevLett.81.5888, PhysRevB.64.155306} and can stabilize the fractional quantum anomalous Hall effect \cite{PhysRevLett.134.196501}, even in an otherwise topologically trivial flat-band system \cite{LIN2026100339}. Interaction-driven orders and quantum geometric effects typically rely on explicit time-reversal symmetry (TRS) breaking or the presence of superconducting order. Their realization more broadly in systems with higher-pseudospin Dirac cones, and whether these Dirac cones govern such phases, are interesting open questions.

\par
\begin{figure}[t]
  \centering
  \centering
    \includegraphics[width=\linewidth]{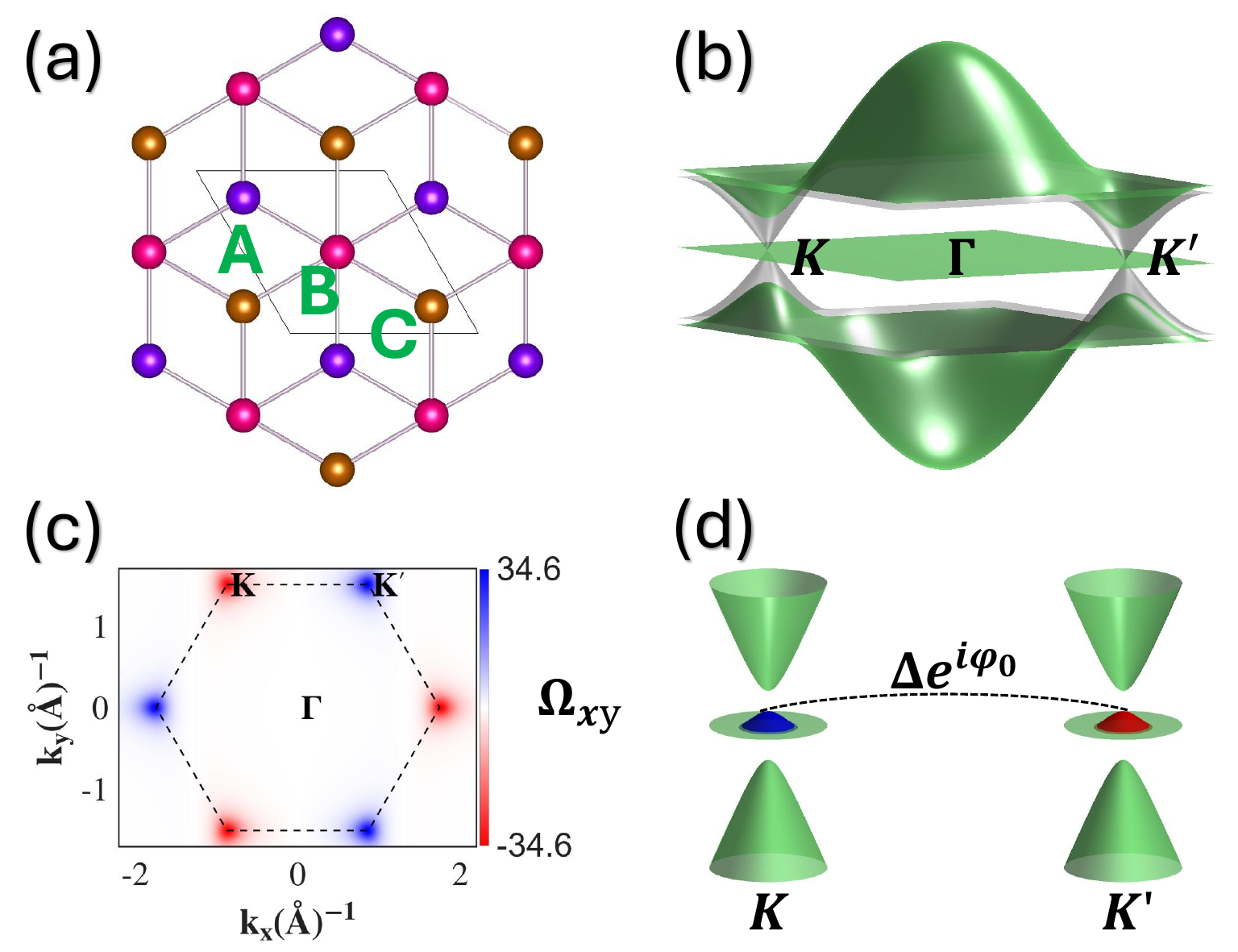}
  \caption{Schematics of (a) the dice lattice and (b) the associated band structures for $m=0$ (gray) and $m\neq0$ (green) within the first BZ. (c) Momentum-space distribution of the flat band's Berry curvature $\Omega_{xy}$ at $m=0.5t$. (d) SVLC order originates via the coherence between the particle (blue) and hole (red) on the flat band of the massive pseudospin-1 Dirac cones.
  }
  \label{fig:01}
\end{figure}

 In this Letter, we address the interaction-driven orders in a partially filled flat band of the dice lattice as an exemplar system. Projecting the interaction onto the isolated flat band reveals a staggered virtual-loop-current (SVLC) order accompanied by a charge order that forms a $\sqrt{3}\times\sqrt{3}$ supercell. These orders arise from the coherence between the pseudospin-1 Dirac cones (Fig.~\ref{fig:01}(d)), revealing an inter-Dirac-cone coherence that goes beyond the conventional IVC. We have established the robustness of our results by reproducing these orders through restricted real-space Hartree-Fock (RSHF) calculations using a tight-binding model with nearest-neighbor (NN) and next-nearest-neighbor (NNN) interactions. Experimental signatures, material realizations, and generalization to other systems are discussed. 

\paragraph*{Dice lattice---} We begin by introducing the tight-binding model for the dice lattice as an illustrative exemplar system, which contains three sublattices (A,B,C) that are NN to each other with hoppings occurring only between the $A$-$B$ and $B$-$C$ sublattices. This preserves $C_{3z}$ symmetry around the axes centered on each sublattice and inversion symmetry centered at the $B$ sublattice (Fig.~\ref{fig:01}(a)). Dice lattice hosts a zero-energy flat band due to its bipartite structure \cite{10.1063/1.1666703, PhysRevB.34.5208, Calugaru2022}. Applying an onsite potential difference between the $A$ and $C$ sublattices breaks the inversion symmetry, and energetically isolates the zero-energy flat band and endows it with a finite Berry curvature. We set the hopping strength to be $t>0$ and the onsite potential difference to $m$. In the $(\phi_A,\phi_B,\phi_C)^T$ basis, the tight-binding Hamiltonian of the dice lattice in momentum space is:
\begin{equation}\label{eq:H_dice}
    H(\mathbf{k})=\begin{pmatrix} m & th(\mathbf{k}) & 0 \\ th^*(\mathbf{k}) & 0 & th(\mathbf{k}) \\ 0 & th^*(\mathbf{k}) & -m \end{pmatrix}.
\end{equation}
Here, $\phi_i$ is a localized orbital on the $i$th sublattice and $h(\mathbf{k})=1+2e^{-i\frac{3k_y}{2}}\cos(\frac{\sqrt{3}k_x}{2})$. The spin degree of freedom is not included as our focus in this study is on orbital effects. We set $\hbar=e=1$ and take the lattice constant $a_d=2.4\,\text{\AA}$ for numerical computations without loss of generality.

\par For $m=0$, the band structure from diagonalizing Eq.~\eqref{eq:H_dice} has pseudospin-1 Dirac cones at the $K$ and $K'$ points in the BZ (Fig.~\ref{fig:01} (b)), as $\lim\limits_{\mathbf{k}\to0}h(\mathbf{k}-\mathbf{K}_\pm)=\pm k_x-ik_y$. Here, $\mathbf{K}_+=\mathbf{K}$ and $\mathbf{K}_-=\mathbf{K}'$ denote the momenta at the $K$ and $K'$ points. The flat band corresponds to the branch of the Dirac cone with zero pseudospin polarization. For $m\neq0$, the Dirac cones become gapped, isolating the flat band at zero energy (Fig.~\ref{fig:01} (b)), and now the Berry curvature concentrates at $K$ and $K'$ points with opposite signs (Fig.~\ref{fig:01} (c)). The strength of the Berry curvature decreases with increasing value of  $m$ as the system approaches the atomic limit. Therefore, we set $m\lesssim t$ to preserve the quantum geometry of the flat band.

\paragraph*{Orbital effect from pseudospin-1 Dirac cones---} Within the Hartree-Fock (HF) approximation, we qualitatively analyze how the pseudospin-1 Dirac cones and Berry curvature stabilize phases driven by short-range interactions.  The pseudospin-1 Dirac cones favor the emergence of complex NNN bond orders underlying the SVLC order, accompanied by a $\sqrt3\times\sqrt3$ charge order, while the Berry curvature acts to bound the strength of these orders.

\par To isolate orbital effects from higher-energy bands, we consider a density-density interaction of strength $V$ in the regime $V\ll m$, such that the interaction does not induce additional band hybridization. The interaction can then be projected onto the flat band, resulting in the Hamiltonian:
\begin{equation}\label{eq:H_int}
    H_{\text{int}} = \int d[\mathbf{r}]d[\mathbf{r'}] V(\mathbf{r}-\mathbf{r'})\hat{n}(\mathbf{r})\hat{n}(\mathbf{r'}),
\end{equation}
where the density is expressed as:
\begin{equation}
    \hat{n}(\mathbf{r}) = \int d[\mathbf{k}]d[\mathbf{k'}] \varphi_{\mathbf{k}}^*(\mathbf{r})\varphi_{\mathbf{k'}}(\mathbf{r})\hat{C}_{\mathbf{k}}^\dagger\hat{C}_{\mathbf{k'}}.
\end{equation}
Here, $V(\mathbf{r}-\mathbf{r'})$ is the interaction potential, $\varphi_{\mathbf{k'}}(\mathbf{r})$ is the flat band's Bloch wavefunction, and $\hat{C}_{\mathbf{k}}$ annihilates a Bloch electron with momentum $\mathbf{k}$. The integral measures are $d[\mathbf{r}]=d^2\mathbf{r}$ and $d[\mathbf{k}]=d^2\mathbf{k}/(2\pi)^2$. Expressing Eq.~\eqref{eq:H_int} in momentum space yields:
\begin{equation}\label{eq:H_int_k}    
    H_{\text{int}} = \int d[\mathbf{q}]d[\mathbf{k}]d[\mathbf{p}] V_{\mathbf{q}}\Lambda(\mathbf{k},\mathbf{q})\Lambda(\mathbf{p},-\mathbf{q})\hat{C}_{\mathbf{k}}^\dagger\hat{C}_{\mathbf{k}-\mathbf{q}}\hat{C}_{\mathbf{p}}^\dagger\hat{C}_{\mathbf{p}+\mathbf{q}},
\end{equation}
where $ V_{\mathbf{q}}$ is the Fourier component of the interaction potential. $\Lambda(\mathbf{k},\mathbf{q})$ is the form factor \cite{Chatterjee2022, PhysRevResearch.5.L012015} defined by
\begin{equation}\label{eq:form_fact_1}
    \Lambda(\mathbf{k},\mathbf{q}) \equiv \int d[\mathbf{r}] e^{i\mathbf{q}\cdot\mathbf{r}}\varphi_{\mathbf{k}}^*(\mathbf{r})\varphi_{\mathbf{k}-\mathbf{q}}(\mathbf{r}) = \braket{u_{\mathbf{k}}|u_{\mathbf{k}-\mathbf{q}}},
\end{equation}
where $u_{\mathbf{k}}(\mathbf{r})=e^{-i\mathbf{k}\cdot\mathbf{r}}\varphi_{\mathbf{k}}(\mathbf{r})$.

\par For short-range interactions and uniform density, taking $\mathbf{q}\to 0$, the Hartree term merely renormalizes the chemical potential, but the Fock term is:
\begin{align}
    H_{\text{Fock}} = & - \int d[\mathbf{q}]d[\mathbf{k}]d[\mathbf{p}] V_{\mathbf{q}}\Lambda(\mathbf{k},\mathbf{q})\Lambda(\mathbf{p},-\mathbf{q}) \notag \\ & \quad \times \left[ \rho_{\mathbf{k},\mathbf{p}+\mathbf{q}} \hat{C}_{\mathbf{p}}^\dagger \hat{C}_{\mathbf{k}-\mathbf{q}} + \text{H.c.} \right], \label{eq:H_Fock}
\end{align}
where $\rho_{\mathbf{k},\mathbf{p}+\mathbf{q}} = \langle \hat{C}_{\mathbf{k}}^\dagger\hat{C}_{\mathbf{p}+\mathbf{q}}\rangle$. A trivial order that preserves all symmetries arises in the case $\rho_{\mathbf{k},\mathbf{p}+\mathbf{q}}\propto\delta_{\mathbf{k},\mathbf{p}+\mathbf{q}}$. A nontrivial coherence channel, however, is found for:
\begin{equation}\label{eq:nontrivial_rho}
\rho_{\mathbf{k},\mathbf{p}+\mathbf{q}} \propto \sum_{\eta=\pm}e^{i\eta\varphi_0} e^{-\frac{\|\mathbf{k}-\mathbf{K}_\eta\|^2}{\Lambda^2}}\delta_{\mathbf{k}+\eta\mathbf{Q},\mathbf{p}+\mathbf{q}} ,
\end{equation}
where $\varphi_0\in[0,2\pi)$, $\mathbf{K}_\pm=\mathbf{K},\mathbf{K}'$, $\mathbf{Q}=\mathbf{K}'-\mathbf{K}$, and $\Lambda\ll m/(t a_d)$. This momentum-space coherence is enabled by the pseudospin-1 Dirac cones, for which $H(\mathbf{K})=H(\mathbf{K}')=\mathrm{diag}(m,0,-m)$, implying the emergence of equivalent flat-band wavefunctions at the $K$ and $K'$ points up to a global phase,
$\braket{u_{\mathbf{K}_\mp}|u_{\mathbf{K}_\pm}}=e^{\pm i\varphi_0}$. Since $\mathbf{k}$ is restricted to $\|\mathbf{k}-\mathbf{K}_\pm\|\ll m/(t a_d)$, where the kinetic energy term remains negligible compared to the mass term, the flat-band wavefunctions remain approximately unchanged from those at $\mathbf{K}$ and $\mathbf{K}'$, yielding
\begin{equation}
\braket{u_{\mathbf{k}\pm\mathbf{Q}}|u_{\mathbf{k}}} \approx e^{\pm i\varphi_0}.
\end{equation}
This naturally favors a coherence channel of the form in Eq.~\eqref{eq:nontrivial_rho}. Consistently, the quantum geometric nestability \cite{PhysRevX.14.041004} shows local minima near the $K$ and $K'$ points; see Supplemental Materials (SM) for details \cite{SM}\nocite{ODA, R_Kubo_1966, RevModPhys.82.1539, PhysRevLett.99.197202}. The Fock term can then be approximated as:
\begin{equation}\label{eq:H_MF}
    H_{\text{Fock}} \approx \sum_{\eta=\pm}\int_{\|\mathbf{k}-\mathbf{K_\eta}\|\ll \frac{m}{ta_d}} d[\mathbf{k}] \left[ \Delta^\eta(\mathbf{k})\hat{C}_{\mathbf{k}+\eta\mathbf{Q}}^\dagger \hat{C}_{\mathbf{k}} + \text{H.c.}\right],
\end{equation}
where the order parameter (OP) is
\begin{equation}\label{eq:OP_0}
    \Delta^\eta(\mathbf{k}) = -\int d[\mathbf{q}] V_{\mathbf{q}}e^{i\eta\varphi_0} |\braket{u_{\mathbf{k}}|u_{\mathbf{k}-\mathbf{q}}}|^2.
\end{equation}
Since short-range interactions probe the $\mathbf{q}\to0$ limit, we expand the form factor in Eq.~\eqref{eq:OP_0} to leading order in $q$, where it is governed by the quantum metric \cite{Chatterjee2022, PhysRevResearch.5.L012015}. This yields:
\begin{equation}\label{eq:OP}
    \Delta^\eta(\mathbf{k}) \approx -V_{\mathbf{0}}e^{i\eta\varphi_0}\int_{\|\mathbf{q}\|\ll\frac{2\pi}{a_d}} d[\mathbf{q}] \left[1-g_{ij}(\mathbf{k})q^iq^j\right],
\end{equation}
where $g_{ij}(\mathbf{k})=\text{Re}[\braket{\partial_{k^i}u_{\mathbf{k}}|(1-\ket{u_{\mathbf{k}}}\bra{u_{\mathbf{k}}})|\partial_{k^j}u_{\mathbf{k}}}]$ is the quantum metric of the flat band \cite{provost1980, PhysRevResearch.7.023158, Yu2025} and $q^i$ is the $i$th component of $\mathbf{q}$. According to Eq.~\eqref{eq:OP}, the magnitude of the OP is upper bounded by the Berry curvature of the flat band, as the latter lower-bounds the trace of the quantum metric \cite{Peotta2015, PhysRevB.90.165139, PhysRevB.95.024515, PhysRevLett.124.167002, PhysRevLett.128.087002, PhysRevX.14.011052}.

\par Based on Eq.~\eqref{eq:H_MF}, the OP establishes a charge order characterized by the wavevector $\mathbf{Q}$, forming a $\sqrt{3}\times\sqrt{3}$ supercell. Since this order originates from states near $K$ and $K'$, where $H(\mathbf{K})=H(\mathbf{K}')=\mathrm{diag}(m,0,-m)$ enforces flat-band wavefunctions localized on the $B$ sublattice, partial filling can be expected to induce inequivalent charge densities and NNN bond orders on the $B$ sublattices. This breaks the $C_{3z}$ symmetry about the axes centered on other sublattices. Moreover, a finite phase $e^{\pm i\varphi_0}$ reflects an emergent $U(1)$ symmetry generated by the conserved relative flat-band occupation at the two pseudospin-1 Dirac cones, $Q=\hat N_{+}-\hat N_{-}$ with $\hat N_{\eta}=\hat C^\dagger_{\mathbf K_\eta}\hat C_{\mathbf K_\eta}$. This emergent symmetry arises since no mechanism provides the large momentum transfer needed to scatter between the two cones. Consequently, the OP spontaneously breaks TRS and exhibits a continuous degeneracy, which is consistent with the evaluation of the energy; see SM for details \cite{SM}.

\begin{figure}[t]
  \centering
  \centering
    \includegraphics[width=\linewidth]{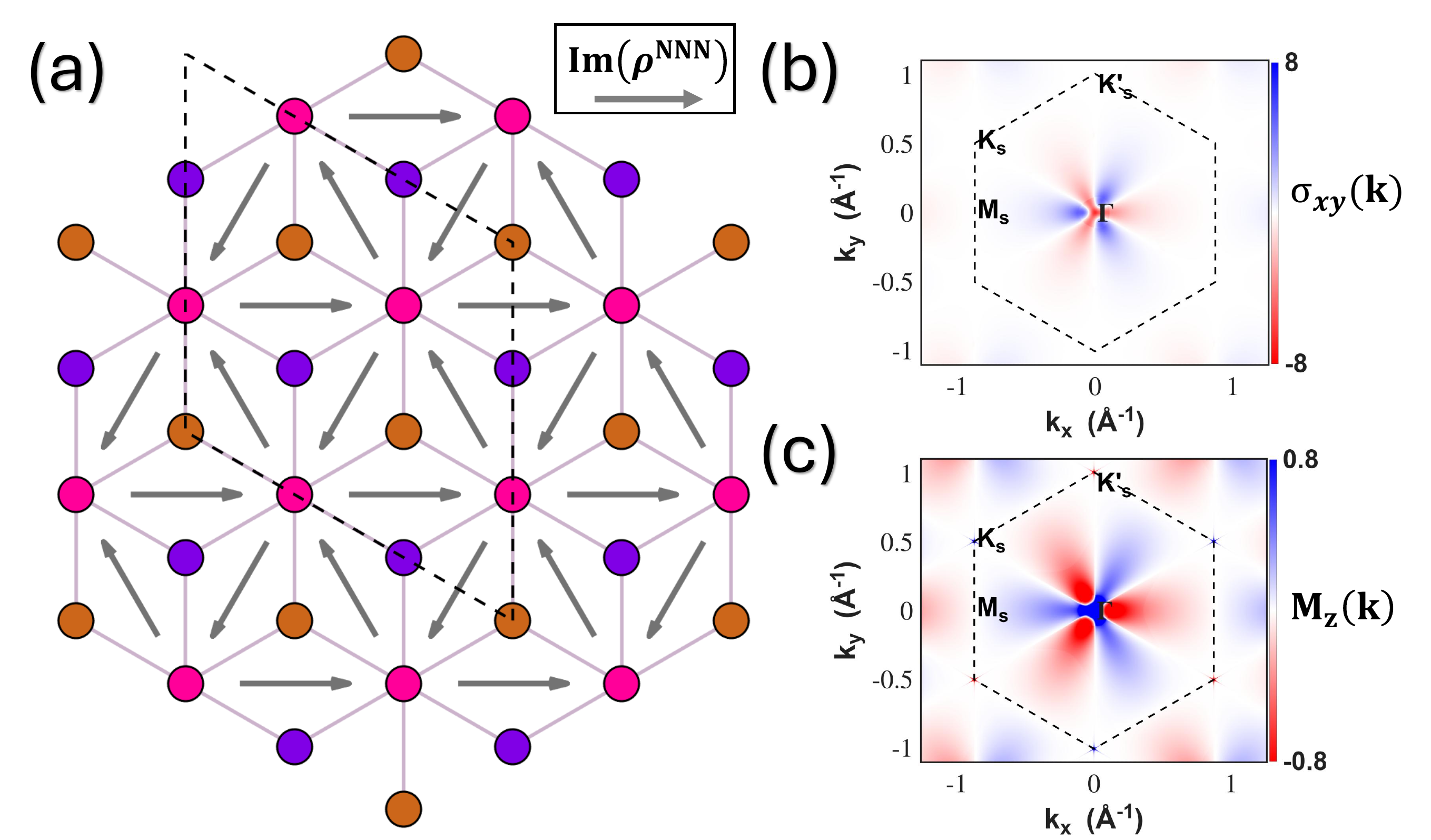}
  \caption{ (a) $\sqrt{3}\times\sqrt{3}$ supercell (black dashed line) of the dice lattice with SVLC order. Gray arrows denote the virtual currents (lengths rescaled). Momentum-space distributions of the (b) intrinsic AHC and (c) OM, with the supercell BZ marked by the black dashed line. 
  }
  \label{fig:02}
\end{figure}

\paragraph*{Staggered virtual-loop-current order---}To demonstrate the analytical insights above, we discuss partially filled flat bands at filling fractions $\nu=1/3$ and $2/3$ in a $\sqrt{3}\times\sqrt{3}$ supercell (Fig.~\ref{fig:02}(a)) using restricted RSHF calculations with NN and NNN interactions of strengths $U_{\text{NN}}$ and $U_{\text{NNN}}$. Motivated by the predicted ordering tendency, we impose the $C_{3z}$ symmetry about the axis centered on the $B$ sublattices to isolate the most robust symmetric orders, while allowing other $C_{3z}$ symmetries to break; see SM for details \cite{SM}.

\par
\begin{figure}[t]
  \centering
  \centering
    \includegraphics[width=\linewidth]{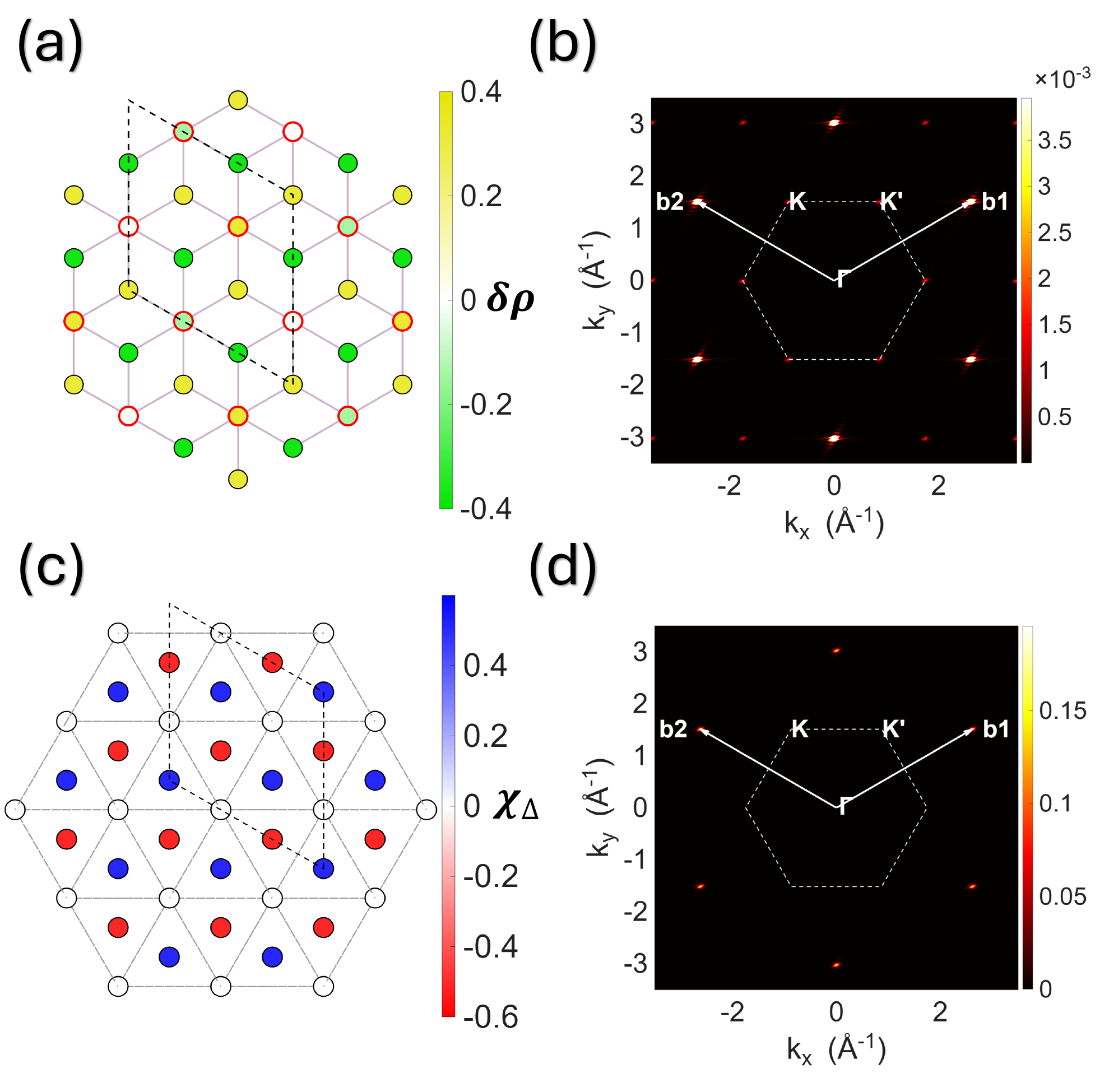}
  \caption{(a) Sublattice-resolved charge density fluctuation $\delta\rho(\mathbf{r})=\rho(\mathbf{r})-\bar{\rho}$ for the charge order, with red circles denoting $B$ sublattices, and (b) the corresponding structure factor $S_{\rho}(\mathbf{q})$. (c) Emergent hexagonal lattice of the VCC $\chi_\triangle$, with empty sites denoting $B$ sublattices, and (d) the corresponding structure factor $S_{\chi_\triangle}(\mathbf{q})$. White dashed lines mark the BZ. $\mathbf{b}_1$ and $\mathbf{b}_2$ are the reciprocal lattice vectors. 
  }
  \label{fig:03}
\end{figure}

 To illustrate the underlying physics, we first present a representative low-lying solution for the parameter values $U_{\text{NN}}=0.05t$, $U_{\text{NNN}}=0.01t$, and $m=0.2t$ at $\nu=1/3$, whose energy per site lies within $6\times10^{-4}t$ above the lowest-energy solution due to the nearly degenerate low-energy manifold. The expected charge and complex NNN bond orders on the $B$ sublattices are more clearly manifest in this solution. The former manifests as density modulations among the inequivalent B sublattices (Fig.~\ref{fig:03}(a)). The latter generates circulating virtual currents around the triangular plaquettes formed by NNN bonds between the B sublattices, with adjacent plaquettes circulating in opposite directions (Fig.~\ref{fig:02}(a)) to form the SVLC order (virtual currents are defined by imaginary parts of the complex NNN bond orders, which break TRS). Due to the absence of single-particle NNN hoppings, the virtual currents are associated with quantum fluctuations of the charge densities \cite{PhysRevB.107.045127} rather than physical charge transport \cite{PhysRevB.110.L041121}. All other NN and NNN bond orders remain real. As a result of spontaneous TRS breaking, a finite intrinsic anomalous Hall conductivity (AHC) $\sigma_{xy}$ (Fig.~\ref{fig:02}(b)) and orbital magnetization (OM) $M_z$ (Fig.~\ref{fig:02}(c)) are generated, reaching $\sigma_{xy}\approx 9.8\times10^{-3}(e^{2}/h)$ and $M_z\approx-6.5\times10^{-3}(t\cdot e/h)$ at a low temperature $k_B T \approx 8.6\times10^{-9}t$. The intrinsic AHC and OM are obtained by constructing the momentum-space mean-field Hamiltonian from the converged RSHF solution; see SM for details \cite{SM}.

\par To characterize the SVLC order, we first define the virtual-current circulation (VCC) $\chi_\triangle$ associated with the $i$th triangular plaquette $\triangle_i$ as
\begin{equation}
\chi_\triangle(\mathbf{R}^\triangle_i)=-\mathrm{Im}\left(\rho_{i_1 i_3}+\rho_{i_3 i_2}+\rho_{i_2 i_1}\right),
\end{equation}
where $(i_1,i_2,i_3)$ label the $B$ sublattices on $\triangle_i$ (going counterclockwise), and $\mathbf{R}^\triangle_i$ denotes the related geometric center. VCC $\chi_\triangle$ forms an emergent hexagonal lattice with alternating signs on the neighboring sites (Fig.~\ref{fig:03}(c)), sharing the unit cell of the dice lattice and reflecting the same degree of TRS breaking. SVLC order is thus characterized by a vanishing average VCC $\bar{\chi}_\triangle=0$ and its nonzero structure factor $S_{\chi_\triangle}(\mathbf{q})$ at the reciprocal lattice vectors of the emergent hexagonal lattice (Fig.~\ref{fig:03}(d)). The charge order is characterized by its nonzero structure factor $S_{\rho}(\mathbf{q})$ at $\mathbf{K}$ and $\mathbf{K'}$ (Fig.~\ref{fig:03}(b)); see SM for details of the structure factors \cite{SM}.

\paragraph{Interaction dependence of the SVLC order---} To gain further insight, we systematically vary strengths of $U_{\text{NN}}$ and $U_{\text{NNN}}$, while keeping them small compared to the mass term ($m=0.2t$). We use $50$ random seeds to search for the lowest-energy solution for each set of parameters.

\begin{figure}[t]
  \centering
  \centering
    \includegraphics[width=\linewidth]{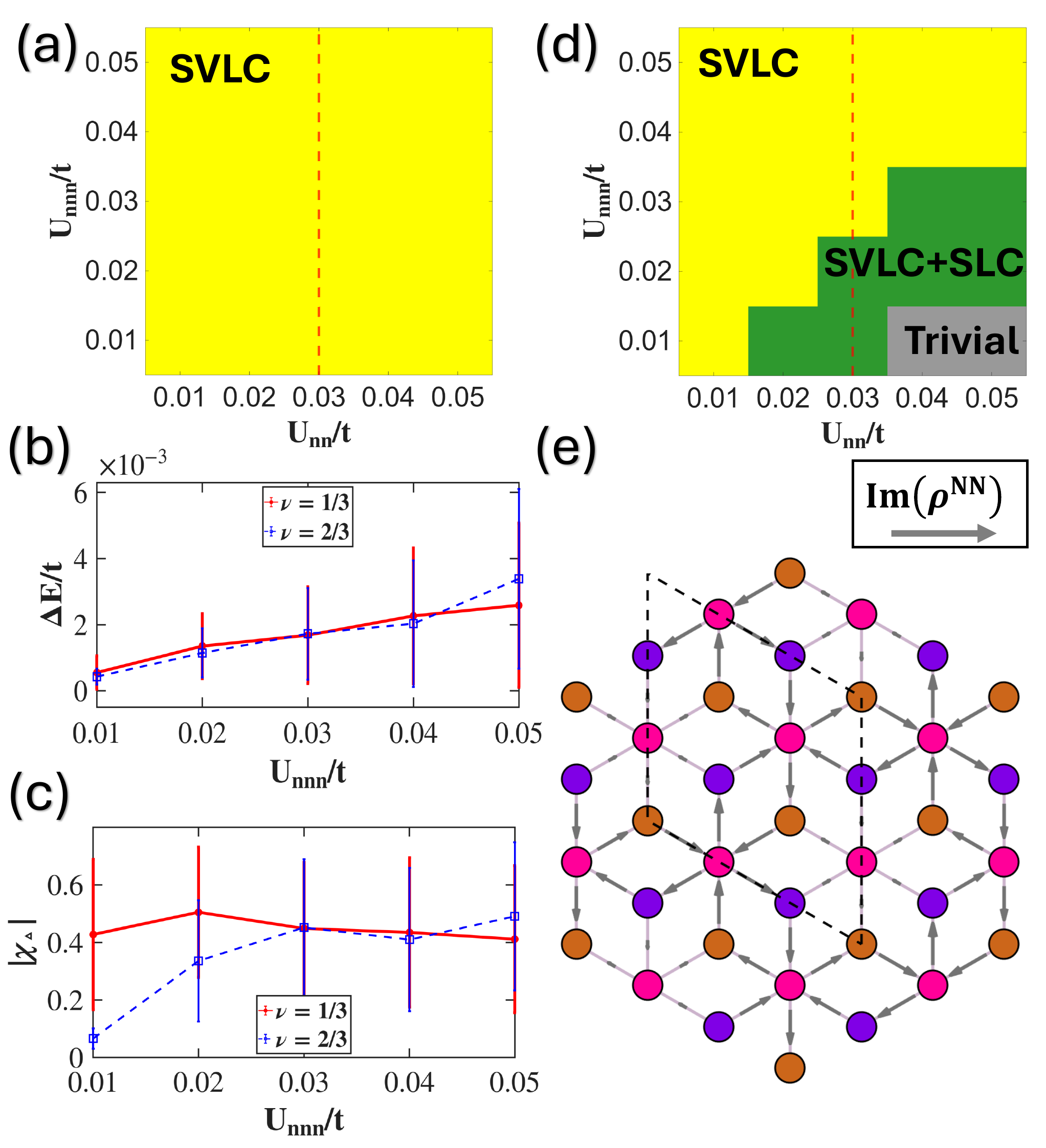}
  \caption{ (a) Phase diagram at $\nu=1/3$. The average and standard deviation of (b) $\Delta E = \langle E_{\text{site}} - \min(E_{\text{site}})\rangle$ and (c) $|\chi_\triangle|$, over random seeds along $U_{\text{NN}}=0.03t$ (red dashed lines in (a,b)). (d) Phase diagram at $\nu=2/3$. (e)  $\sqrt{3}\times\sqrt{3}$ supercell (black dashed line) of the dice lattice with the SLC order. Gray arrows denote the currents (lengths rescaled).
  }
  \label{fig:04}
\end{figure}

\par Throughout the weakly interacting regime at $\nu=1/3$, all solutions exhibit the SVLC order (Fig.~\ref{fig:04}(a)) with broad variations in $|\chi_\triangle|$, with $\sigma_{|\chi_\triangle|}\sim|\chi_\triangle|_{\mathrm{avg}}$. For small $U{\mathrm{NNN}}$ values, the solutions are nearly degenerate, with $\Delta E,\sigma_E\ll U_{\mathrm{NN}},U_{\mathrm{NNN}}$. Representative examples are shown in Figs.~\ref{fig:04}(b) and (c); see SM for the full map \cite{SM}. Here, $\Delta E=\langle E_{\mathrm{site}}-\min(E_{\mathrm{site}})\rangle$ and $|\chi_\triangle|_{\text{avg}}$ denote the seed-averaged relative energy per site with respect to the lowest-energy solution and the seed-averaged $|\chi_\triangle|$, while $\sigma_E$ and $\sigma_{|\chi_\triangle|}$ are the corresponding standard deviations, respectively. This near degeneracy qualitatively reflects the emergent $U(1)$ symmetry, while its weak lifting in the lattice-model calculations may arise from corrections beyond the ideal flat-band-projected theory, such as nontrivial Hartree contributions from inequivalent sublattice connectivities. Although the near degeneracy becomes progressively less pronounced as interactions become longer-ranged with increasing $U_{\mathrm{NNN}}$, all solutions continue to exhibit SVLC order.

\par Although the predicted order remains robust at $\nu=2/3$, we find an additional coexisting staggered loop-current (SLC) order, which is not present at $\nu=1/3$ (Fig.~\ref{fig:04}(d)), corresponding to loop currents around the parallelogram plaquettes formed by the two B sites and their shared A/C sites, with neighboring plaquette circulating in opposite directions (Fig.~\ref{fig:04}(e)); see SM for details \cite{SM}. In contrast to the SVLC order, such loop currents arise from complex NN bond orders, which carry physical currents and induce local magnetic moments that may be accessible to spin-polarized neutron diffraction \cite{PhysRevB.110.195109} and tuning fork resonator experiments \cite{Gui2025}, as explored in vanadium-based kagome metals \cite{Fernandes2026}. Notably, within the parameter range we considered, the lowest energy solution at $\nu=2/3$ exhibits a trivial order that preserves all symmetries in a low-$U_{\text{NNN}}/U_{\text{NN}}$ regime (Fig.~\ref{fig:04}(d)). In contrast, this trivial order is unstable at $\nu=1/3$; see SM for details \cite{SM}. These results indicate that $\nu=2/3$ filling hosts a rich phase structure \cite{footnote}. 

\paragraph{Experimental signatures and material realization---}In the absence of interaction-driven orders, the dice lattice preserves TRS and therefore exhibits no intrinsic AHC and OM. Upon emergence of the SVLC and charge orders, TRS and translational symmetry are spontaneously broken. The former produces finite intrinsic AHC and OM measurable via transport and magnetic probes, respectively, while the latter would be detectable via scanning tunneling microscopy.

\par Several candidate materials and heterostructures have previously been proposed to realize the dice lattice, including YCl \cite{geng2025YCldice}, SrTiO$_3$/SriIrO$_3$/SriTiO$_3$ superlattices \cite{doi:10.7566/JPSJ.87.041006}, LaAlO$_3$/SriTiO$_3$ (111) quantum wells \cite{PhysRevLett.111.126804}, Hg$_{\rm 1-x}$Cd$_{\rm x}$Te \cite{PhysRevB.92.035118}, TMO (111) trilayer \cite{Xiao2011}, metallic networks \cite{PhysRevLett.83.5102, PhysRevLett.86.5104} and CO molecules on Cu(111) surfaces \cite{Tassi_2024, Gomes2012, Khajetoorians2019}. The required onsite potential difference here could be achieved via atomic substitution or an out-of-plane electric field in multilayer heterostructures, offering a promising direction for experimental realization. Beyond solid-state materials, ultracold atoms in optical lattices could provide a synthetic platform to realize the dice lattice \cite{PhysRevA.80.063603, PhysRevLett.108.045306, PhysRevA.92.033617}, where the anomalous Hall effect associated with the SVLC order can be probed via transverse deflection measurements \cite{Aidelsburger2015}.

\paragraph{Discussion---} While our proposed SVLC order and IVC both involve coherence between Dirac cones in the BZ, IVC specifically requires pseudospin-1/2 Dirac cones to constitute the only low-energy states near specific BZ momentum points, thereby defining a valley degree of freedom. In contrast, the zero-energy flat band in the dice lattice invalidates such a low-energy-valley picture. Instead, the pseudospin-1 Dirac cones enter the SVLC order through the equivalence of their zero-energy flat-band wavefunctions. Consequently, the pseudospin-1 Dirac cones remain the dominant low-energy degrees of freedom despite the additional zero-energy states away from them. This reveals a fundamental distinction between the underlying mechanism of spontaneous symmetry breaking phases in pseudospin-1/2 and pseudospin-1 Dirac systems, where Dirac cones are no longer synonymous with valleys.

\par Beyond the illustrative example of the dice lattice, the same physics extends to the doubled honeycomb lattice \cite{Neves2024}, where one sublattice is duplicated and vertically separated without interconnection while remaining connected to the other sublattice. This yields an identical band structure to the dice lattice. Consequently, a similar SVLC order is expected in the doubled honeycomb lattice, suggesting that the flat-band-wavefunction-based mechanism proposed here will likely extend to a broader class of pseudospin-1 Dirac systems.

\par Our results demonstrate how the interplay between Dirac-cone physics and flat-band wavefunctions can give rise to emergent ordered states beyond the established valley-based descriptions. These findings open new directions for exploring interaction-driven phenomena in pseudospin-1 Dirac systems.

\paragraph*{Acknowledgment---}
We are grateful to Professor Robert Markiewicz for important discussions in connection with Hartree-Fock computations. This work was supported by the National Science Foundation through the Expand-QISE award NSF-OMA-2329067 and benefited from the resources of Northeastern University’s Advanced Scientific Computation Center, the Explorer Cluster, the Massachusetts Technology Collaborative award MTC-22032, and the Quantum Materials and Sensing Institute (QMSI).

\paragraph*{Data availability---}The data that support the findings of this article are not publicly available upon publication because it is not technically feasible and/or the cost of preparing, depositing, and hosting the data would be prohibitive within the terms of this research project. The data are available from the authors upon reasonable request.



\bibliography{apssamp}
\setcounter{equation}{0}
\setcounter{figure}{0}
\setcounter{table}{0}

\renewcommand{\theequation}{S\arabic{equation}}
\renewcommand{\thefigure}{S\arabic{figure}}
\renewcommand{\thetable}{S\arabic{table}}
\renewcommand{\bibnumfmt}[1]{[S#1]}
\renewcommand{\citenumfont}[1]{S#1}
\newcommand{\bk}{\boldsymbol\kappa}

\newcommand{\beginsupplement}{%
  \setcounter{equation}{0}
  \renewcommand{\theequation}{S\arabic{equation}}%
  \setcounter{table}{0}
  \renewcommand{\thetable}{S\arabic{table}}%
  \setcounter{figure}{0}
  \renewcommand{\thefigure}{S\arabic{figure}}%
  \setcounter{section}{0}
  \renewcommand{\thesection}{S\Roman{section}}%
  \setcounter{subsection}{0}
  \renewcommand{\thesubsection}{S\Roman{section}.\Alph{subsection}}%
}

\clearpage
\pagebreak
\begin{widetext}
\begin{center}
\textbf{\large Supplemental Materials: Inter-Dirac-Cone Coherence Beyond the Valley description}
\end{center}
\tableofcontents

\section{S1. Quantum-geometric nestability of the flat band}
\par We discuss the quantum geometric nestability $\omega_0^{(\mathbf{Q})}$ of the flat band in the dice lattice as a function of nesting vector $\mathbf{Q}$. We focus on the particle-hole channel in which $\omega_0^{(\mathbf{Q})}$ is characterized by the lowest eigenvalue of the following linear operator \cite{PhysRevX.14.041004}:
\begin{equation}
    \Pi_{\mu'\nu';\mu\nu}^{(\mathbf{Q})} = \frac{\Omega_{\text{BZ}}}{(2\pi)^2} \sum_{\mathbf{k}}
    P_{\mu'\mu}\!\left(\mathbf{k}+\frac{\mathbf{Q}}{2}\right)
    Q_{\nu\nu'}\!\left(\mathbf{k}-\frac{\mathbf{Q}}{2}\right)
    +
    Q_{\mu'\mu}\!\left(\mathbf{k}+\frac{\mathbf{Q}}{2}\right)
    P_{\nu\nu'}\!\left(\mathbf{k}-\frac{\mathbf{Q}}{2}\right).
\end{equation}
Here, $\Omega_{\text{BZ}}$ is the area of BZ, $\mu,\nu$ label basis orbitals, $P$ is the projection operator onto the flat band, and $Q=1-P$. $\omega_0^{(\mathbf{Q})}$ quantifies the deviation from perfect quantum geometric nesting (QGN), in which $\omega_0^{(\mathbf{Q})}=0$ and there exists a nesting matrix $N^{(\mathbf{Q})}$ satisfying $\sum_{\mu\nu}\Pi_{\mu'\nu';\mu\nu}^{(\mathbf{Q})}N_{\mu\nu}^{(\mathbf{Q})}=0$. This implies that the quantum geometry of the flat band favors an order parameter of the form $[O^{(p-h)}]_{mn}\sim \frac{\Omega_{\text{BZ}}}{(2\pi)^2} \sum_{\mathbf{k}}\left[U^\dagger(\mathbf{k}+\mathbf{Q}/2)N^{(\mathbf{Q})}U(\mathbf{k}-\mathbf{Q}/2)\right]_{mn}$ \cite{PhysRevX.14.041004}, where $m,n$ denotes the flat-band indices and $U(\mathbf{k})$ relates to $P(\mathbf{k})$ by $P(\mathbf{k})=U(\mathbf{k})U^\dagger(\mathbf{k})$. 

\par In our case, one may focus on the flat band near the pseudospin-1 Dirac cones, where the Hamiltonian reads
\begin{equation}
    \lim_{\mathbf{k}\to0}H(\mathbf{k}+\mathbf{K}_\pm) \approx \begin{pmatrix} m &  t(\pm k_x-ik_y) & 0 \\ t(\pm k_x+ik_y) & 0 & t(\pm k_x-ik_y) \\ 0 & t(\pm k_x+ik_y) & -m \end{pmatrix},
\end{equation}
with the corresponding flat-band wavefunction
\begin{equation}
    \ket{\psi_0(\mathbf{k}+\mathbf{K}_\pm)} \propto \begin{pmatrix} -\frac{t}{m}(\pm k_x-ik_y) \\ 1 \\ \frac{t}{m}(\pm k_x+ik_y) \end{pmatrix}.
\end{equation}
Here, $\mathbf{K}_+=\mathbf{K}$ and $\mathbf{K}_-=\mathbf{K}'$ indicate the momenta at the $K$ and $K'$ points. Note that since
\begin{equation}
    \ket{\psi_0(\mathbf{k}+\mathbf{K}_-)} =
    \begin{pmatrix}
        0 & 0 & 1 \\
        0 & 1 & 0 \\
        1 & 0 & 0
    \end{pmatrix}
    \ket{\psi_0(\mathbf{k}+\mathbf{K}_+)} \equiv N^{(\mathbf Q)}\ket{\psi_0(\mathbf{k}+\mathbf{K}_+)},
\end{equation}
the flat-band wavefunctions locally exhibit ``perfect nesting" with $\mathbf Q=\pm(\mathbf K_- - \mathbf K_+)$ near the two pseudospin-1 Dirac cones. Although this does not persist over the full BZ and hence does not constitute perfect QGN in the strict sense, it heuristically suggests relatively better quantum geometric nestability near these $\mathbf Q$.

\par We have verified the preceding qualitative analysis via numerical computation of $\omega_0^{(\mathbf{Q})}$ of the flat band in the dice lattice as a function of $\mathbf{Q}$ (Fig.~\ref{fig:QGN}). The global minimum occurs at $\mathbf Q=0$ and yields a perfect QGN. However, this case admits only the trivial nesting matrix, $N^{(\mathbf Q)}=\mathbb{1}$, and does not support a nontrivial coherence channel. Instead, as shown in Fig.~\ref{fig:QGN}, $\omega_0^{(\mathbf Q)}$ develops local minima at all symmetry-equivalent nesting vectors $\mathbf Q=\pm(\mathbf K_- - \mathbf K_+)$, with values of order $10^{-2}$. This supports the nontrivial coherence channel discussed in the main text, as it indicates relatively better quantum geometric nestability features. It would be interesting to investigate the nesting matrix near $\mathbf Q=\pm(\mathbf K_- - \mathbf K_+)$, the classes of interactions that may stabilize the corresponding coherence channels, and other local minima of $\omega_0^{(\mathbf Q)}$ in future studies.

\begin{figure}[h]
  \centering
  \centering
    \includegraphics[width=\linewidth]{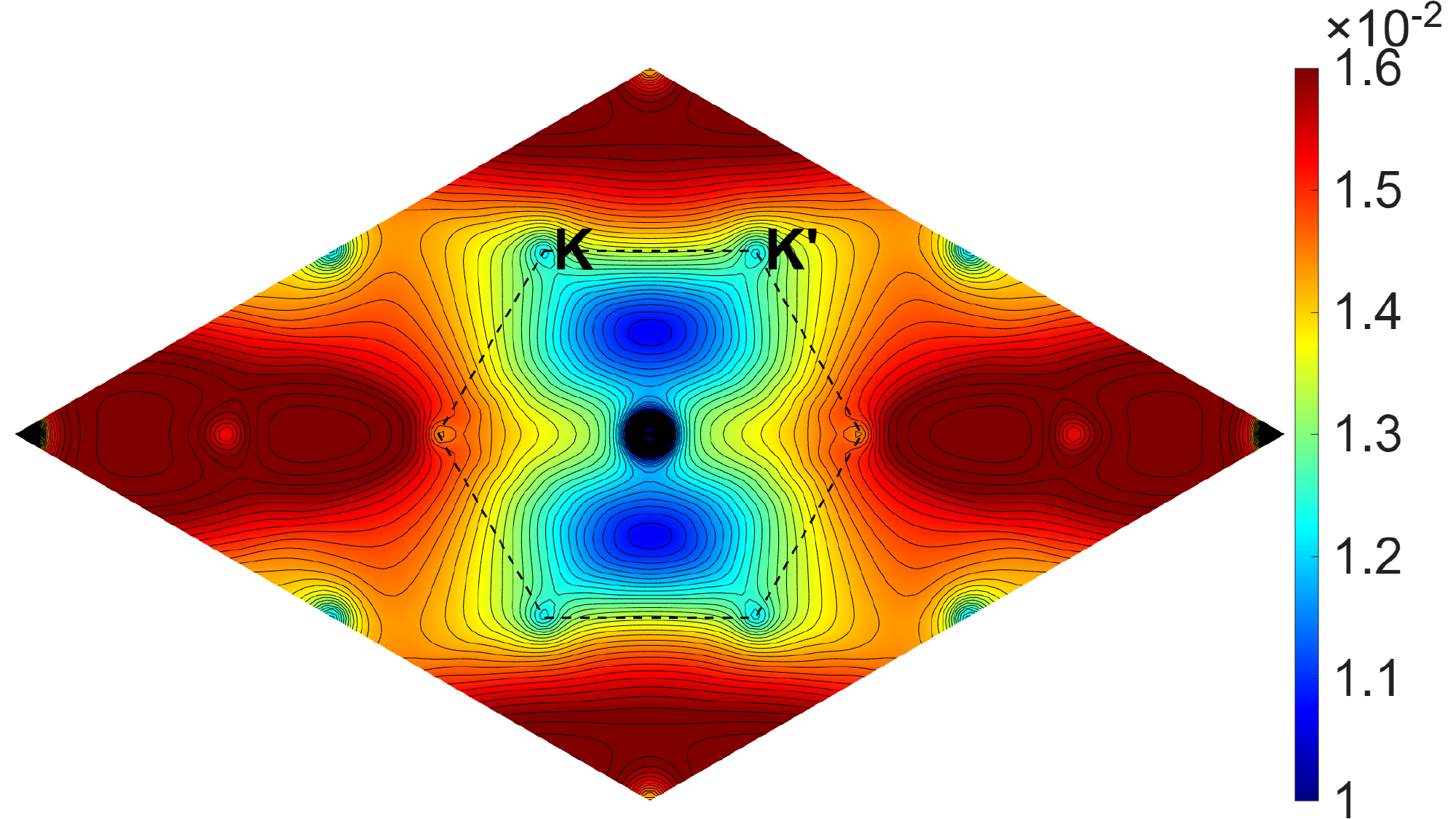}
  \caption{Quantum geometric nestability $\omega_0^{(\mathbf{Q})}$ of the flat band in the dice lattice as a function of $\mathbf{Q}$. Black dashed line marks the unit-cell BZ, and the black solid contours denote constant-$\omega_0^{(\mathbf{Q})}$ lines to help visualize the local minima.
  }
  \label{fig:QGN}
\end{figure}

\section{S2. Analytical evaluation of the Hartree-Fock energy}
\par We analytically evaluate the energy to demonstrate the continuous degeneracy introduced by the phase $e^{\pm i\varphi_0}$ of the OP in Eq.~(11) of the main text. Since the Hartree term only renormalizes the chemical potential under the assumptions of uniform density and short-range interactions, we only need to consider the Fock contribution for evaluating the energy. Recall that the Fock term has the form:
\begin{align}
    H_{\text{Fock}} = & - \int d[\mathbf{q}]d[\mathbf{k}]d[\mathbf{p}] V_{\mathbf{q}}\Lambda(\mathbf{k},\mathbf{q})\Lambda(\mathbf{p},-\mathbf{q})   \left[ \rho_{\mathbf{k},\mathbf{p}+\mathbf{q}} \hat{C}_{\mathbf{p}}^\dagger \hat{C}_{\mathbf{k}-\mathbf{q}} + \text{H.c.} \right],
\end{align}
where $\Lambda(\mathbf{k},\mathbf{q}) = \braket{u_{\mathbf{k}}|u_{\mathbf{k}-\mathbf{q}}}$. This yields the energy as:
\begin{equation}\label{eq:E_Fock}
    E_{\text{Fock}} = -\frac{1}{2} \int d[\mathbf{q}]d[\mathbf{k}]d[\mathbf{p}] V_{\mathbf{q}}\braket{u_{\mathbf{k}}|u_{\mathbf{k}-\mathbf{q}}}\braket{u_{\mathbf{p}}|u_{\mathbf{p}+\mathbf{q}}}\left[ \rho_{\mathbf{k},\mathbf{p}+\mathbf{q}} \rho_{\mathbf{p},\mathbf{k}-\mathbf{q}} + \text{c.c.} \right].
\end{equation}
The focused nontrivial channel is:
\begin{equation}\label{eq:OP_ansatz}
\rho_{\mathbf{k},\mathbf{p}+\mathbf{q}} \propto  \sum_{\eta=\pm}e^{i\eta\varphi_0} e^{-\frac{\|\mathbf{k}-\mathbf{K}_\eta\|^2}{\Lambda^2}}\delta_{\mathbf{k}+\eta\mathbf{Q},\mathbf{p}+\mathbf{q}},
\end{equation}
where $\varphi_0\in[0,2\pi)$, $\mathbf{K}_\pm=\mathbf{K},\mathbf{K}'$, $\mathbf{Q}=\mathbf{K}'-\mathbf{K}$, and $\Lambda\ll m/(t a_d)$. By combining Eqs.~\eqref{eq:E_Fock} and \eqref{eq:OP_ansatz} under short-range interactions where the $\mathbf{q}\to0$ limit is considered, we get:
\begin{equation}\label{eq:E_Fock_3}
    E_{\text{Fock}} \cong - \int_{\mathbf{q}\ll\frac{2\pi}{a_d}} d[\mathbf{q}] \sum_{\eta=\pm}\int_{\|\mathbf{k}-\mathbf{K}_\eta\|\ll\frac{m}{t a_d}} d[\mathbf{k}] V_{\mathbf{q}}\braket{u_{\mathbf{k}}|u_{\mathbf{k}-\mathbf{q}}}\braket{u_{\mathbf{k}+\eta\mathbf{Q}-\mathbf{q}}|u_{\mathbf{k}+\eta\mathbf{Q}}}.
\end{equation}
Since $E_{\text{Fock}}$ is independent of the phase $e^{\pm i\varphi_0}$, the continuous degeneracy is justified.

\section{S3. Details of the restricted RSHF calculations}
\par We discuss the variational parameters used in our restricted RSHF calculations, along with the relevant computational details. In an RSHF calculation, the mean-field Hamiltonian is:
\begin{equation}\label{eq:H_MF_0}
    H_{\text{MF}} = H_{\text{TB}} + H_{\text{Hartree}}[\rho] + H_{\text{Fock}}[\rho],
\end{equation}
With the tight-binding Hamiltonian:
\begin{align}
    H_{\text{TB}} = t\sum_{\langle i,j \rangle}\left[\hat{C}^\dagger_{A,i}\hat{C}_{B,j} + \hat{C}^\dagger_{B,i}\hat{C}_{C,j}  + \text{H.c.} \right] + m\sum_{i}\hat{C}^\dagger_{A,i}\hat{C}_{A,i} - \hat{C}^\dagger_{C,i}\hat{C}_{C,i}. \label{eq:H_TB_SM}
\end{align}
Upon applying interactions between all NN and NNN sites, the Hartree and Fock terms are:
\begin{align}
H_{\text{Hartree}}[\rho] = & \sum_{\alpha,\beta}\bigg[ U_{\text{NN}}\sum_{\langle i,j\rangle}\rho_{\beta j\beta j}\hat{C}^\dagger_{\alpha,i}\hat{C}_{\alpha,i}  + U_{\text{NNN}}\sum_{\langle\langle i,j\rangle\rangle}\rho_{\beta j\beta j}\hat{C}^\dagger_{\alpha,i}\hat{C}_{\alpha,i} \bigg], \label{eq:H_Hatree_SM}
\\ H_{\text{Fock}}[\rho] = & -\sum_{\alpha,\beta}\bigg[ U_{\text{NN}}\sum_{\langle i,j\rangle}\rho_{\beta j\alpha i}\hat{C}^\dagger_{\alpha,i}\hat{C}_{\beta,j} + U_{\text{NNN}}\sum_{\langle\langle i,j\rangle\rangle}\rho_{\beta j\alpha i}\hat{C}^\dagger_{\alpha,i}\hat{C}_{\beta,j} \bigg]. \label{eq:H_Fock_SM}
\end{align}
Here, $\hat{C}_{\alpha,i}$ annihilates an electron on the $i$th site with $\alpha$th type of sublattice, where $\alpha=A,B,C$. $\langle ... \rangle$ indicates the NN sites and $\langle\langle ... \rangle\rangle$ indicates the NNN sites. The variational parameters are defined as
\begin{equation}
\rho_{\beta j\alpha i}=\langle \hat{C}_{\beta,j}^\dagger\hat{C}_{\alpha,i}\rangle,
\end{equation}
where the expectation value is taken over the $N_{\text{occ.}}$ eigenstates of $H_{\text{MF}}$ with the lowest energies \cite{PhysRevB.110.L041121}. The occupation number $N_{\text{occ.}}$ is determined by
\begin{equation}
    N_{\text{occ.}} = \frac{N_{\text{site}}^{\sqrt{3}\times\sqrt{3}}}{N_{\text{site}}^{1\times1}}(1+\nu),
\end{equation}
where $N_{\text{site}}^{\sqrt{3}\times\sqrt{3}}$ and $N_{\text{site}}^{1\times1}$ are the number of lattice sites in the $\sqrt{3}\times\sqrt{3}$ supercell and unit cell, respectively. $\nu=1/3$ or $2/3$ is the filling fraction of the flat band. The energy per site is calculated via
\begin{equation}    
    E_{\text{site}}[\rho] = \frac{1}{2N_{\text{site}}^{\sqrt{3}\times\sqrt{3}}}\text{Tr}[\rho(H_{\text{TB}}+H_{\text{MF}})].
\end{equation}

\par Our initial unrestricted RSHF calculations revealed a rugged energy landscape with many nearly degenerate, seed-dependent solutions, as expected for an isolated exactly flat band due to the absence of kinetic energy. In this case, the lowest-energy solutions exhibit glassy states that break all symmetries and give little information about the Dirac cone physics of interest. Therefore, we performed the RSHF calculations restricted by the $C_{3z}$ symmetry, as already noted in the main text. Specifically, the restriction preserves the $C_{3z}$ symmetry about the axis on the $B$ sublattices but allows other $C_{3z}$ symmetries to be broken. The resulting charge densities are uniform within the $A$ and $C$ sublattices, with identical NN bond orders between the $A$ and $C$ sublattices as well as the NNN bond orders within the $A$ and $C$ sublattices. In contrast, orders involving the $B$ sublattices are not constrained to be uniform. Therefore, the variational parameters reduce to five types of charge densities
\begin{align}
\rho_A = & \sum_{i=1}^3\frac{\rho_{A i A i}}{3}, 
\\ \rho_C = & \sum_{i=1}^3\frac{\rho_{C i C i}}{3},
\\ \rho_{B_i} = & \rho_{B_i i B_i i},
\end{align}
seven types of NN bond orders 
\begin{align}
\rho_{AB_i} = & \sum_{j\in\langle i,j \rangle} \frac{\rho_{B_i i A j}}{3},
\\ \rho_{B_iC} = & \sum_{j\in\langle i,j \rangle} \frac{\rho_{B_i i C j}}{3},
\\ \rho_{AC} = & \sum_{\langle i,j \rangle} \frac{\rho_{A i C j}}{9},
\end{align}
and five types of NNN bond orders
\begin{align}
\rho_{AA}^{(\text{NNN})} = & \sum_{\langle\langle i,j \rangle\rangle} \frac{\rho_{A_i i A_j j}}{3},
\\ \rho_{B_iB_j}^{(\text{NNN})} = & \rho_{B_i i B_j j},
\\ \rho_{CC}^{(\text{NNN})} = & \sum_{\langle\langle i,j \rangle\rangle} \frac{\rho_{C_i i C_j j}}{3}.
\end{align}
We applied periodic boundary conditions in iteratively updating the NN and NNN bond orders during the self-consistency cycles.

\begin{figure}[h]
  \centering
  \centering
    \includegraphics[width=\linewidth]{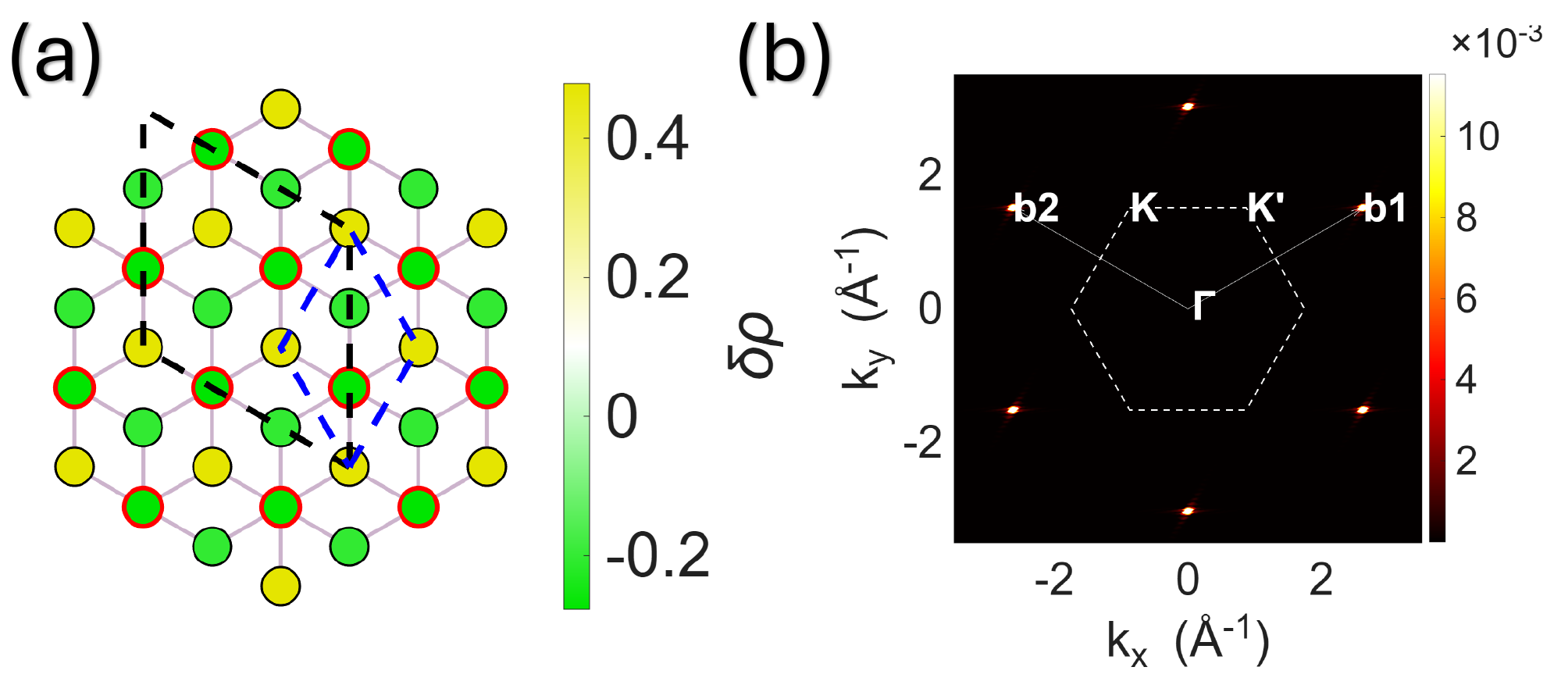}
  \caption{ (a) Charge density fluctuations of the trivial order. Red circles mark $B$ sublattices in the $\sqrt{3}\times\sqrt{3}$ supercell. Black and blue dashed lines indicate the $\sqrt{3}\times\sqrt{3}$ supercell and the unit cell, respectively. (b) The corresponding structure factor $S_\rho(\mathbf{q})$, where the white dashed line marks the unit-cell BZ. $\mathbf{b}_1$ and $\mathbf{b}_2$ are the reciprocal lattice vectors.
  }
  \label{fig:S1}
\end{figure}

\par During self-consistency cycles, we applied the optimal damping algorithm \cite{ODA} to help accelerate convergence. The convergence criteria required that the change in energy per site and the values of the variational parameters between successive iterations fall below $10^{-15}t$ and $10^{-15}$, respectively. 

\par Upon further restricting the variational parameters for the NN bond orders by using  $(\rho_{AB}=\sum_{i=1}^{3}\rho_{AB_i}/3,\rho_{BC}=\sum_{i=1}^{3}\rho_{B_iC}/3,\rho_{AC})$, the trivial order that preserves all symmetries (Fig.~\ref{fig:S1}) becomes metastable at $\nu=1/3$ with energy per site around $0.11t$ above the SVLC order. Here, computations were performed at $U_{\text{NN}}=0.05t$, $U_{\text{NNN}}=0.01t$, and $m=0.2t$. Nevertheless, the latter remains the lowest-energy solution, whereas the former becomes unstable once the restriction is relaxed. We verified such an instability at each point of the interaction scan.

\section{S4. Momentum-space mean-field Hamiltonian}
\par After obtaining $\rho_{\beta j\alpha i}$ from a converged RSHF computation, we can construct the corresponding mean-field Hamiltonian via Eq.~\eqref{eq:H_MF_0}, which can be straightforwardly transformed into the momentum space to compute the band structure (Fig.~\ref{fig:BS}) as well as the intrinsic contribution of anomalous Hall conductivity (AHC) and orbital magnetization (OM). In Fig.~2 (b,c) of the main text and Fig.~\ref{fig:BS}, the Fermi level $\mu_0$ is determined by searching for $\mu$ that satisfies $\sum_{n}\int d[\mathbf{k}]f_n(\mathbf{k},\mu)=N_{\text{occ.}}$, where $f_n(\mathbf{k},\mu)$ is the Fermi-Dirac distribution function of the $n$th HF band with chemical potential $\mu$. The temperature is set at $k_BT\approx 8.6\times10^{-9}t$. Results shown in Fig.~2 (b,c) of the main text are based on a $600\times600$ $k$-point mesh; increasing the mesh size was found to yield no significant changes.

\begin{figure}[h]
  \centering
    \includegraphics[width=0.6\linewidth]{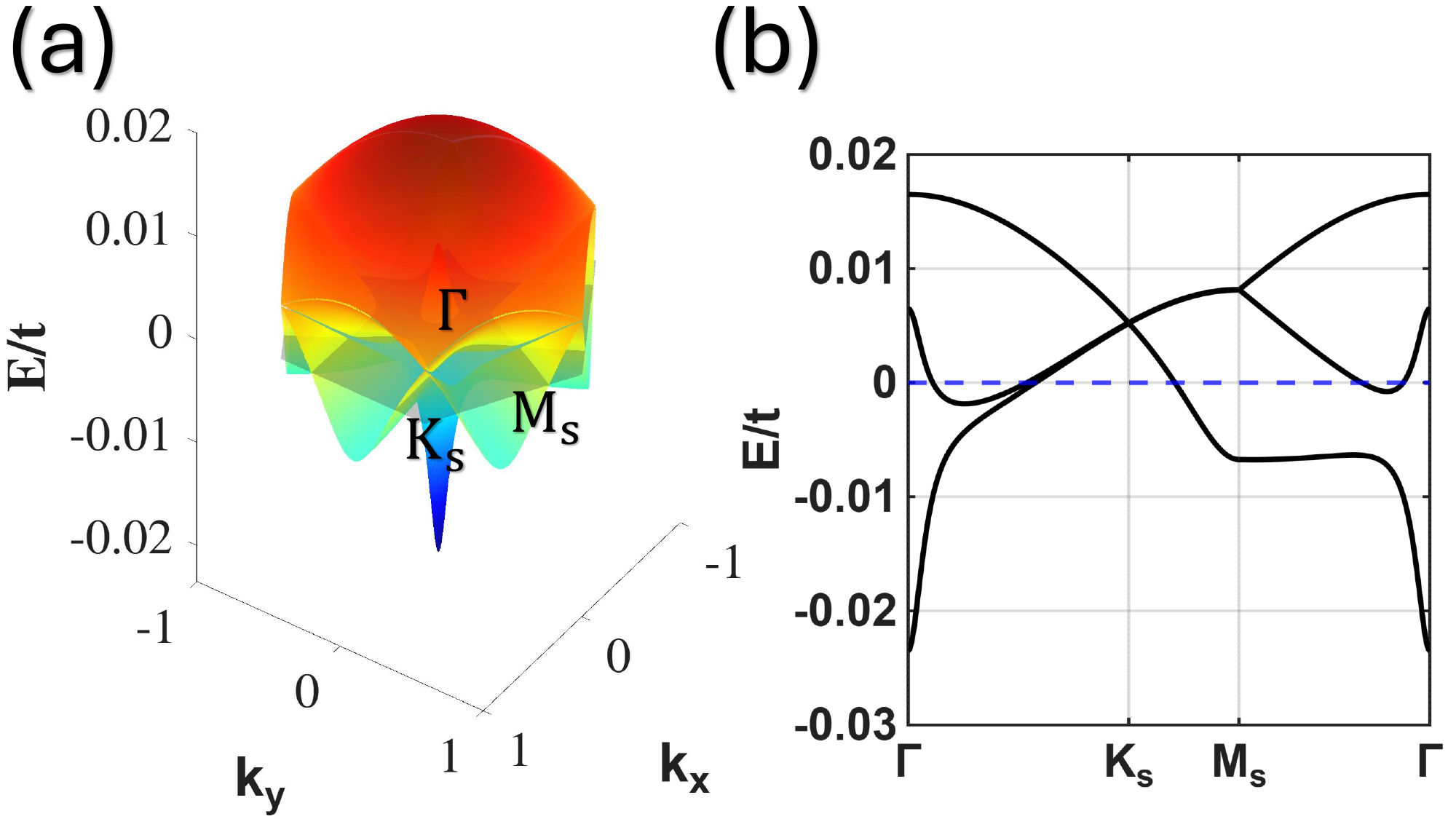}
  \caption{ HF band structure of the flat bands (a) over the supercell BZ and (b) along a high-symmetry path with the Fermi level set to zero, computed from the representative low-lying solution at $U_{\text{NN}}=0.05t$, $U_{\text{NNN}}=0.01t$, and $m=0.2t$ at $\nu=1/3$. The gray plane in (a) and the blue dashed line in (b) mark the Fermi level $\mu_0$.
  } 
  \label{fig:BS}
\end{figure}

\par The intrinsic contribution of the AHC was calculated via the Kubo formula \cite{R_Kubo_1966, RevModPhys.82.1539}:
\begin{equation}
    \sigma_{xy} = e^2\hbar\int d[\mathbf{k}] \sigma_{xy}(\mathbf{k}),
\end{equation}
where 
\begin{align}
    \sigma_{xy}(\mathbf{k}) & = \sum_{n}f_n(\mathbf{k},\mu_0)\Omega_{n,xy}(\mathbf{k}),
    \\ \Omega_{n,xy}(\mathbf{k}) & = -\sum_{m\neq n}\text{Im}\left[\frac{(v^x_{\mathbf{k}})_{nm}(v^y_{\mathbf{k}})_{mn}}{(E_m(\mathbf{k})-E_n(\mathbf{k}))^2}\right],
    \\ (v^\mu_{\mathbf{k}})_{nm} & = \frac{1}{\hbar}\braket{\tilde{u}_{n\mathbf{k}}|\frac{\partial H_{\text{MF}}(\mathbf{k})}{\partial k_\mu}|\tilde{u}_{m\mathbf{k}}}.
\end{align}
Here, $H_{\text{MF}}(\mathbf{k})$ is the mean-field Hamiltonian in momentum space, and $E_m(\mathbf{k})$ and $\tilde{u}_{m\mathbf{k}}(\mathbf{r})$ are the energy and the cell-periodic Bloch wavefunction of the $m$th band, respectively. 

\par The OM is calculated via \cite{PhysRevLett.99.197202}:
\begin{equation}
    M_z = \frac{e}{\hbar}\int d[\mathbf{k}] M_z(\mathbf{k}),
\end{equation}
where 
\begin{align}
    M_z(\mathbf{k}) & = \sum_{n} f_n(\mathbf{k},\mu_0)m_n^{(z)}(\mathbf{k}) + k_BT\ln(1+e^{-\frac{E_{n}(\mathbf{k})-\mu_0}{k_BT}})\Omega_{n,xy}(\mathbf{k}), \label{eq:M_tot}
    \\ m_n^{(z)}(\mathbf{k}) & = -\text{Im}\left[ \braket{\partial_{k^x} \tilde{u}_{n\mathbf{k}}| \left( E_{n}(\mathbf{k}) - H_{\text{MF}}(\mathbf{k})\right) |\partial_{k^y}\tilde{u}_{n\mathbf{k}}} \right].
\end{align}
To avoid numerical instabilities from differentiation and low-temperature divergences, we reformulate $m_n^{(z)}(\mathbf{k})$ and approximate the logarithmic function in the second term of Eq.~\eqref{eq:M_tot} as:
\begin{align}
    m_n^{(z)}(\mathbf{k}) & = -\hbar^2\sum_{m\neq n}\text{Im}\left[ \frac{(v^x_{\mathbf{k}})_{nm}(v^y_{\mathbf{k}})_{mn}}{E_{n}(\mathbf{k})-E_{m}(\mathbf{k})} \right],
    \\ \lim_{k_BT\to0}k_BT\ln(1+e^{-\frac{E_{n}(\mathbf{k})-\mu_0}{k_BT}}) & \approx -(E_{n}(\mathbf{k})-\mu_0)f_n(\mathbf{k},\mu_0).
\end{align}

\section{S5. Definitions of structure factors}
\par Here we define the structure factors that are used to characterize the SVLC and charge orders. The structure factor for characterizing the SVLC order is defined as:
\begin{align}\label{eq:struct_fact_LC-like}
S_{\chi_\triangle}(\mathbf{q}) & = \left|\sum_i \left[ \chi_\triangle(\mathbf{R}^\triangle_i) - \bar{\chi}_\triangle \right] e^{-i\mathbf{q}\cdot\mathbf{R}^\triangle_i}\right|^2.
\end{align}
The charge density structure factor is:
\begin{align}
S_{\rho}(\mathbf{q}) & = \left|\sum_i \left[ \rho(\mathbf{r}_i)-\bar{\rho} \right] e^{-i\mathbf{q}\cdot\mathbf{r}_i}\right|^2,
\end{align}
where $\rho(\mathbf{r}_i)$ is the charge density at the $i$th site located at $\mathbf{r}_i$, and $\bar{\rho}$ is the average charge density.

\section{S6. Statistics related to the interaction scan}
\par We present the average and standard deviation of the energy per site measured relative to the lowest-energy solution $\Delta E= \langle E_{\text{site}} - \text{min}(E_{\text{site})}\rangle$ (Fig.~\ref{fig:S2}) and VCC strength $|\chi_\triangle|$ (Fig.~\ref{fig:S3}). Note that the SVLC order persists in regimes with larger interaction strengths, where $\Delta E$ and its standard deviation $\sigma_E$ become comparable to the interaction strength. This feature is more pronounced in regimes where $U_{\text{NNN}}\geq U_{\text{NN}}$. This regime, however, is generally not expected for short-range repulsive interactions, where the interaction strength typically decreases with distance. Therefore, the gradual lifting of the near degeneracy is expected.

\begin{figure}[h]
   \centering
   \centering
     \includegraphics[width=\linewidth]{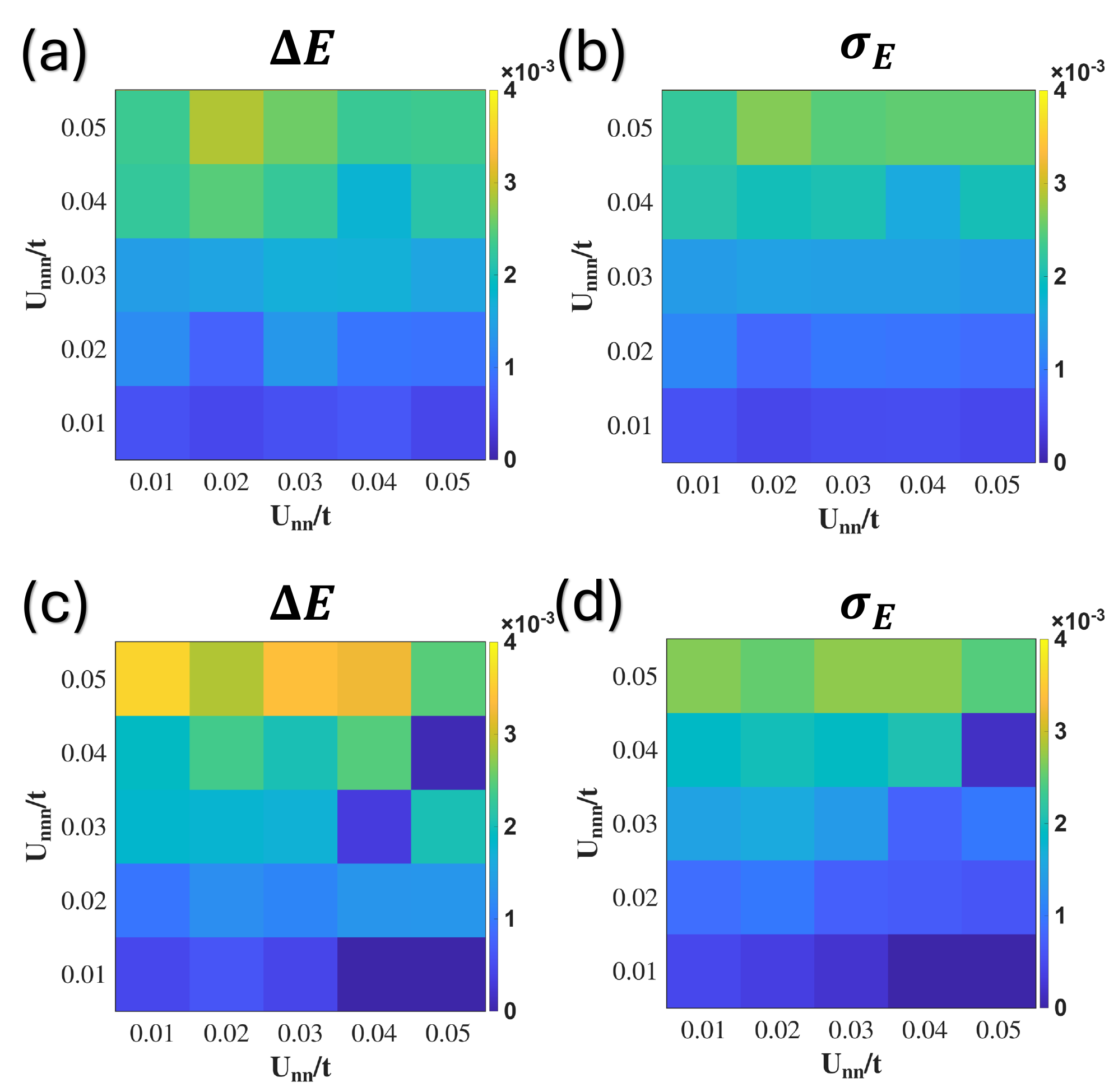}
   \caption{ (a) Average $\Delta E= \langle E_{\text{site}} - \text{min}(E_{\text{site})}\rangle$ and (b) standard deviation $\sigma_E$ of the energy per site measured relative to the lowest-energy solution over random seeds at $\nu=1/3$, and the (c) average and (d) standard deviation at $\nu=2/3$.
   }
   \label{fig:S2}
\end{figure}

\begin{figure}[h]
   \centering
   \centering
     \includegraphics[width=\linewidth]{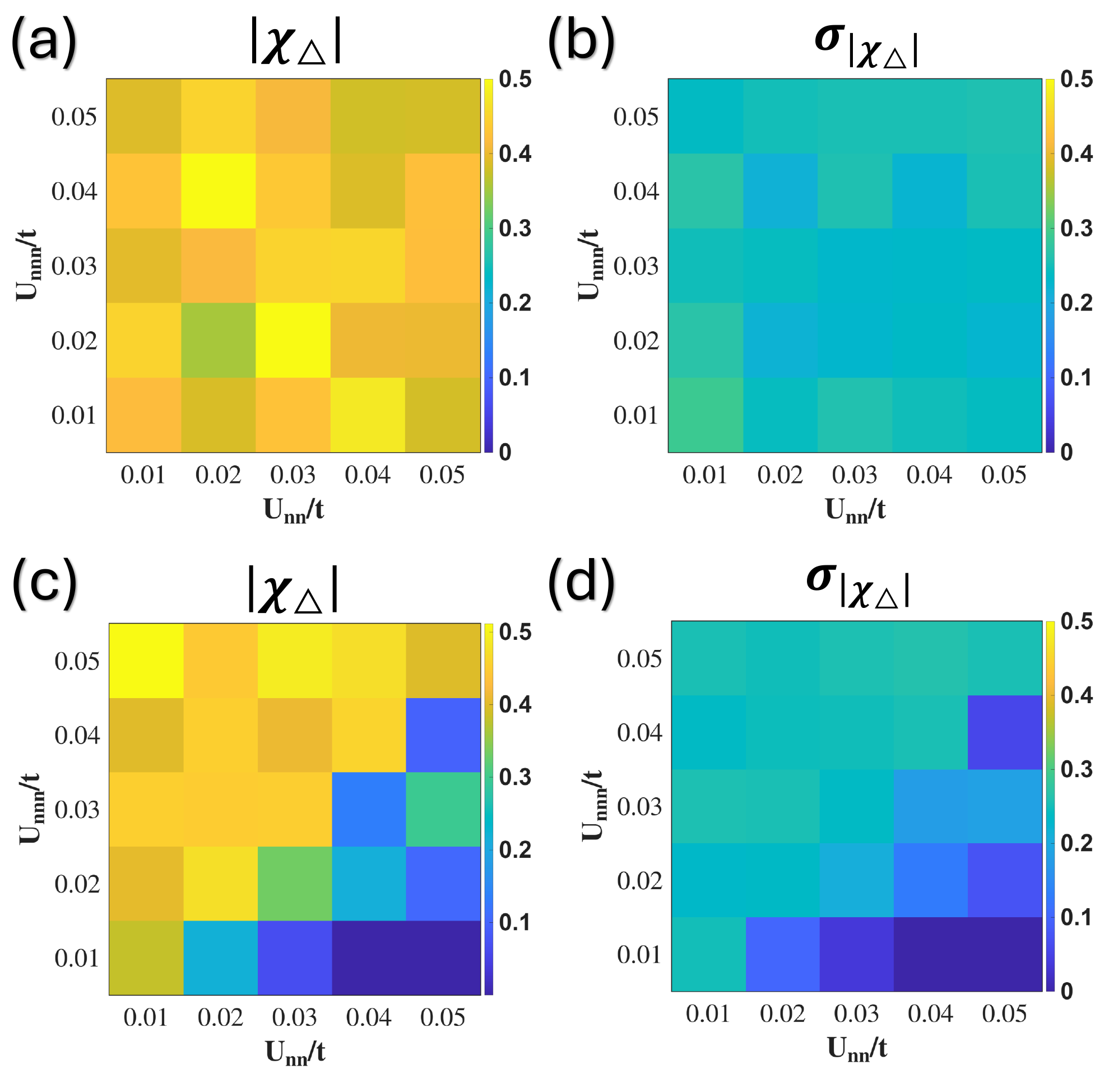}
   \caption{(a) Average $|\chi_\triangle|_{\text{avg}}$ and (b) standard deviation $\sigma_{|\chi_\triangle|}$ of the VCC strength $|\chi_\triangle|$ over random seeds at $\nu=1/3$, and the (c) average and (d) standard deviation at $\nu=2/3$.
   }
   \label{fig:S3}
\end{figure}

\section{S7. Characterizing the staggered loop-current order}
\par To characterize the SLC order, we define the associated dimensionless magnetic moment $m_{\lozenge}(\mathbf{R}^\lozenge_i)$ on the $i$th parallelogram plaquette:
\begin{equation}
    m_{\lozenge}(\mathbf{R}^\lozenge_i) = -\mathrm{Im}\left(\rho_{i_1 i_4} + \rho_{i_4 i_3}+\rho_{i_3 i_2}+\rho_{i_2 i_1}\right),
\end{equation}
where $(i_1,i_2,i_3,i_4)$ label the sublattices on $\lozenge_i$ taken going counterclockwise, and $\mathbf{R}^\lozenge_i$ denotes its geometric center. Since $m_{\lozenge}$ forms an emergent kagome lattice (Fig.~\ref{fig:S4}(a)) within the $\sqrt{3}\times\sqrt{3}$ supercell, the SLC order is characterized by the peak of its structure factor at $\mathbf{K}$ and $\mathbf{K'}$ (Fig.~\ref{fig:S4}(b)), which is defined as:
\begin{align}\label{eq:struct_fact_LC-like}
S_{m_{\lozenge}}(\mathbf{q}) & = \left|\sum_i \left[ m_{\lozenge}(\mathbf{R}^\lozenge_i) - \bar{m}_\lozenge \right] e^{-i\mathbf{q}\cdot\mathbf{R}^\lozenge_i}\right|^2,
\end{align}
where $\bar{m}_\lozenge$ is the average dimensionless magnetic moment.

\par Since the physical magnetic moment $M_{\lozenge_i}^{(\text{phys})}$ induced by the loop currents on the $i$th plaquette is:
\begin{equation}
    m_{\lozenge_i}^{(\text{phys})} = \int_{\lozenge_i} \mathbf{r}\times Jd\mathbf{l} = \sum_{l,n\in\lozenge_i} -2\frac{te}{\hbar}\text{Im}[\rho_{ln}]\cdot\frac{a_d^2}{4\sqrt{3}} = \frac{tea_d^2}{2\sqrt{3}\hbar}m_{\lozenge}(\mathbf{R}^\lozenge_i),
\end{equation}
$m_{\lozenge}(\mathbf{R}^\lozenge_i)$ serves as a faithful proxy for the observable local magnetic moment.

\begin{figure}[h]
   \centering
   \centering
     \includegraphics[width=\linewidth]{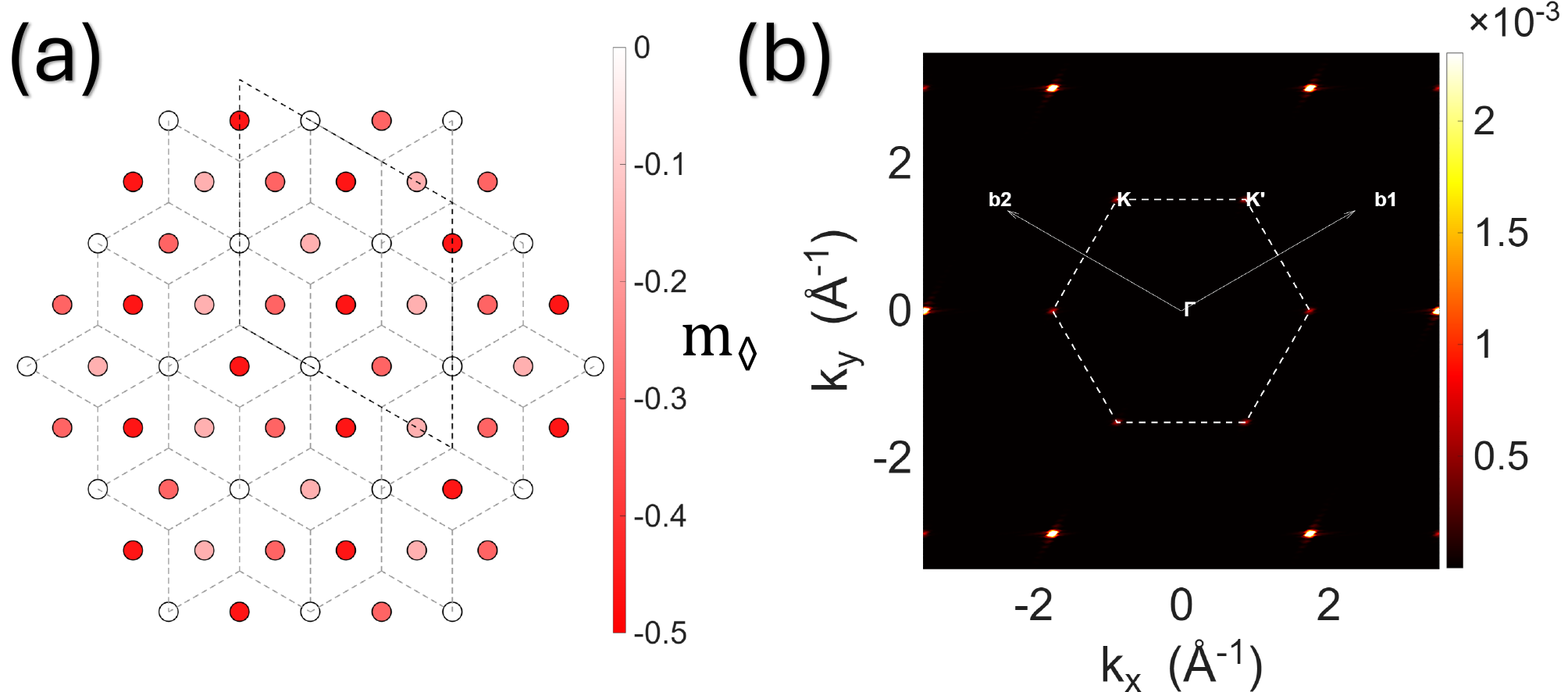}
   \caption{(a) The emergent kagome lattice of the magnetic moment $m_{\lozenge}$, where empty sites denote $B$ sublattices, dashed gray lines NN bonds, and the dashed black line the $\sqrt{3}\times\sqrt{3}$ supercell. (b) The corresponding structure factor $S_{m_{\lozenge}}(\mathbf{q})$. White dashed lines mark the unit-cell BZ. $\mathbf{b}_1$ and $\mathbf{b}_2$ are the reciprocal lattice vectors. 
   }
   \label{fig:S4}
\end{figure}

\end{widetext}
\end{document}